\documentclass[sigconf,authorversion]{acmart}

\usepackage[utf8]{inputenc}
\usepackage{enumitem}
\usepackage{amsmath}
\usepackage{nicefrac}

\usepackage{algorithm}
\usepackage[noend]{algpseudocode}
\algrenewcommand{\alglinenumber}[1]{\color{gray}\tiny#1:}
\algnewcommand{\algorithmicglobals}{\textbf{globals}}
\algdef{SxnE}[GLOBALS]{Globals}{EndGlobals}
   {\algorithmicglobals}
\algrenewcommand{\Call}[2]{\textbf{#1}(#2)}
\algnewcommand{\Fail}[1]{\Call{fail}{\enquote{#1}}}
\makeatletter
\algnewcommand{\LeftComment}[2]{%
\hspace*{#1\ALG@thistlm}\(\triangleright\) #2}
\makeatother
\makeatletter
\newcommand{\StateInd}[1][3]{%
  \setlength\@tempdima{\algorithmicindent}%
  \Statex\hskip\dimexpr#1\@tempdima\relax}
\makeatother

\usepackage{xspace}

\usepackage[nohyperlinks,nolist]{acronym}
\newcommand{\gls}{\ac}

\begin{acronym}[HWASan]
\acro{aslr}[ASLR]{Address Space Layout Randomization}
\acro{ir}[IR]{Intermediate Representation}
\acro{asan}[ASan]{AddressSanitizer}
\acro{hwasan}[HWASan]{Hardware-assisted AddressSanitizer}
\acro{cwe}[CWE]{Common Weakness Enumeration}
\acro{cve}[CVE]{Common Vulnerabilities and Exposures}
\acro{cfi}[CFI]{Control-Flow Integrity}
\acro{lto}[LTO]{Link Time Optimization}
\end{acronym}

\newcommand{\Name}{MESH\xspace}
\newcommand{\Cpp}{C\texttt{++}\xspace}
\newcommand{\archx}{x86-64\xspace}
\newcommand{\archarm}{ARM64\xspace}
\newcommand*{\eg}{e.g.,\@\xspace}
\newcommand*{\ie}{i.e.,\@\xspace}

\usepackage{pifont}% http://ctan.org/pkg/pifont

\newcommand{\goalone}{\ding{182}\xspace}
\newcommand{\goaltwo}{\ding{183}\xspace}
\newcommand{\goalthree}{\ding{184}\xspace}
\newcommand{\goalfour}{\ding{185}\xspace}
\newcommand{\goalfive}{\ding{186}\xspace}
\newcommand{\goalsix}{\ding{187}\xspace}

\usepackage{cleveref}

\usepackage{csquotes}
\copyrightyear{2021}
\acmYear{2021}
\setcopyright{acmlicensed}\acmConference[ARES 2021]{The 16th International Conference on Availability, Reliability and Security}{August 17--20, 2021}{Vienna, Austria}
\acmBooktitle{The 16th International Conference on Availability, Reliability and Security (ARES 2021), August 17--20, 2021, Vienna, Austria}
\acmPrice{15.00}
\acmDOI{10.1145/3465481.3465760}
\acmISBN{978-1-4503-9051-4/21/08}


\begin{document}

%
% The "title" command has an optional parameter, allowing the author to define a "short title" to be used in page headers.
\title{\Name: A Memory-Efficient Safe Heap for C/\Cpp}

\begin{abstract}
While memory corruption bugs stemming from the use of unsafe programming languages are
an old and well-researched problem, the resulting vulnerabilities still
dominate real-world exploitation today.
Various mitigations have been proposed to alleviate the problem, mainly in the
form of language dialects, static program analysis, and code or binary
instrumentation. Solutions like AdressSanitizer (ASan) and Softbound/CETS have proven that
the latter approach is very promising, being able to
achieve memory safety without requiring manual source code adaptions,
albeit suffering substantial performance and memory overheads.
While performance overhead can be seen as a flexible constraint, extensive memory
overheads can be prohibitive for the use of such solutions in
memory-constrained environments.
To address this problem, we propose \Name, a highly memory-efficient safe
heap for C/\Cpp. 
With its constant, very small memory overhead (configurable up to $2$~MB on \archx) and constant
complexity for pointer access checking, \Name offers efficient, byte-precise
spatial and temporal memory safety for memory-constrained scenarios.
Without jeopardizing the security of safe heap objects,
\Name is fully compatible with existing code and uninstrumented libraries,
making it practical to use in heterogeneous environments.
We show the feasibility of our approach with a full LLVM-based prototype supporting both major
architectures, \ie \archx and \archarm, in a Linux runtime environment.
Our prototype evaluation shows that, compared to ASan and
Softbound/CETS, \Name can achieve huge memory savings while
preserving similar execution performance.

\end{abstract}

%\begin{CCSXML}
%<ccs2012>
%  <concept_id>10010520.10010553.10010562</concept_id>
%  <concept_desc>Computer systems organization~Embedded systems</concept_desc>
%  <concept_significance>500</concept_significance>
%  <concept_id>10010520.10010575.10010755</concept_id>
%  <concept_desc>Computer systems organization~Redundancy</concept_desc>
%  <concept_significance>300</concept_significance>
%  <concept_id>10010520.10010553.10010554</concept_id>
%  <concept_desc>Computer systems organization~Robotics</concept_desc>
%  <concept_significance>100</concept_significance>
%  <concept_id>10003033.10003083.10003095</concept_id>
%  <concept_desc>Networks~Network reliability</concept_desc>
%  <concept_significance>100</concept_significance>
%</ccs2012>
%\end{CCSXML}
%
%\ccsdesc[500]{Computer systems organization~Embedded systems}

%
% The "author" command and its associated commands are used to define the authors and their affiliations.
% Of note is the shared affiliation of the first two authors, and the "authornote" and "authornotemark" commands
% used to denote shared contribution to the research.
\author{Emanuel Q. Vintila}
\email{emanuel.vintila@gmail.com}
\orcid{}
\affiliation{%
  \institution{Fraunhofer AISEC}
  \streetaddress{}
  \city{Garching, near Munich}
  \state{}
  \postcode{}
  \country{Germany}
}
\author{Philipp Zieris}
\email{philipp.zieris@aisec.fraunhofer.de}
\orcid{0000-0001-9658-1572}
\affiliation{%
  \institution{Fraunhofer AISEC}
  \streetaddress{}
  \city{Garching, near Munich}
  \state{}
  \postcode{}
  \country{Germany}
}
\author{Julian Horsch}
\email{julian.horsch@aisec.fraunhofer.de}
\orcid{0000-0001-9018-7048}
\affiliation{%
  \institution{Fraunhofer AISEC}
  \streetaddress{}
  \city{Garching, near Munich}
  \state{}
  \postcode{}
  \country{Germany}
} 

%
% By default, the full list of authors will be used in the page headers. Often, this list is too long, and will overlap
% other information printed in the page headers. This command allows the author to define a more concise list
% of authors' names for this purpose.
%\renewcommand{\shortauthors}{Vintila, et al.}

%
% This command processes the author and affiliation and title information and builds
% the first part of the formatted document.
\maketitle

\newpage
\section{Introduction}
The C and \Cpp programming languages are preferred for system programming
because they are efficient and offer low-level control.
These advantages come at the cost of memory safety, requiring the programmer
to manually manage heap memory and ensure the legality of memory accesses. 
Incorrect handling by the programmer 
can lead to serious issues~\cite{SPWS13} such as
\emph{buffer overflows} or \emph{use-after-free} bugs, which can be exploited
by attackers to hijack the control-flow (causing the program to inadvertently
change its course of 
execution)~\cite{Sha07,BRSS08,CDD+10,STL+15}
or to leak information~\cite{SPP+04,GGK+16}.
Even though memory vulnerabilities are long-known issues, they are
still very common in mainstream programs: Out of all the vulnerabilities
reported in the \gls{cve} database, 14.7\% are marked as overflows and 4.3\%
are marked as memory corruptions \cite{CVE}. 
C/\Cpp memory corruption bugs can be divided into two categories:
\begin{description}
\item[Temporal memory bugs] access an object outside of its lifetime, \ie
	before its \emph{allocation} or after its \emph{deallocation}.
\item[Spatial memory bugs] 
	access an object outside of its bounds, \ie using addresses \emph{lower than
	the start} or \emph{higher than the end} of the object.
\end{description}
Several mechanisms have been proposed to
mitigate the effects of memory bugs. 
A widely used mitigation mechanism are stack
canaries~\cite{CPM+98}, detecting overflows on the stack before they can affect
the control flow via the stack-based return address.
Another mitigation mechanism is \gls{aslr}
\cite{LHBF14},
randomizing the location of program parts, making it harder
for the attacker to correctly guess valid code addresses. 
Finally, in recent years, a plethora of \gls{cfi} approaches have been proposed
\cite{BCN+17,BZP19},
trying to protect indirect jumps, calls, and function returns using
various policies and techniques.
Unfortunately, trading security for performance typically leaves
mitigation techniques vulnerable to exploitation, as shown in various attacks on
\gls{cfi}~\cite{GABP14,CCD+15,CBP+15} and
\gls{aslr}~\cite{SPP+04,BBM+14,RSB+17}.

A more complete way to counter memory vulnerabilities is to detect and mitigate the
memory bugs themselves---and not just their effects---by imposing \emph{memory safety}.
Several ways of enforcing memory safety in C/\Cpp have been proposed,
among which are \emph{language dialects},
\emph{static analysis}, and, more relevant to our work, \emph{code or binary instrumentation}.
Language dialects remove unsafe features from the languages, offering
alternatives, for example, in the form of new pointers 
types~\cite{ABS94,JMG+02,NMW02}. 
While they can be effective 
and efficient (fewer run-time checks),
dialects of C/\Cpp lack compatibility with existing code and are less likely to be accepted in
practice. 
Static analysis techniques \cite{JH11,BCD+18} detect bugs without running
the code, and hence do not impose any run-time overheads, but are less
effective. 
Finally, code or binary instrumentation can
be used for detecting memory bugs through dynamic analysis
\cite{SWS+16,MHH+19,SLR+19}.
They work on existing code and allow the programmer to write new code as usual,
without any knowledge about the underlying instrumentation. 

The most prominent instrumentation-based memory safety mechanism is \gls{asan}~\cite{SBPV12}.
\gls{asan} approximates memory safety by inserting protected memory regions
between objects and delaying reuse of freed memory.
A more precise solution is offered by SoftBound/CETS~\cite{NZMZ09,NZMZ10},
which ties pointers to their objects and achieves strong temporal and spatial
memory safety. Both approaches, as well as other similar
solutions~\cite{BMCP18,DY16,SSS+18}, 
impose substantial run-time and memory
overheads. 
While a large performance overhead might be
impractical, it presents a flexible constraint and does not 
prevent a solution from being applicable at all. In contrast, extensive memory
overheads can be prohibitive for applying a solution in memory-constrained
devices. 
Furthermore, most of the approaches~\cite{ABS94,JMG+02,NMW02,BMCP18,DY16,NZMZ09,NZMZ10} do
not offer a concept for (secure) compatibility with uninstrumented code, 
hindering their real-world applicability, \eg in situations where external
libraries cannot be instrumented.

To tackle those problems, we present \Name, a novel, highly memory-efficient 
heap safety mechanism, which uses a configurable but constant-sized lookup table, the
\textit{\Name table}, for storing
bounds and the validity of objects in memory.
\Name makes use of the unused bits in a pointer (by default, on Linux, only 47 bits of a
pointer are used on \archx\footnote{\url{https://www.kernel.org/doc/html/v5.8/x86/x86_64/mm.html}},
and 48 on \archarm\footnote{\url{https://www.kernel.org/doc/html/v5.8/arm64/memory.html}})
to link pointers to their objects' metadata entries in the \Name table.
Using a constant lookup algorithm, \Name ensures that
only legal pointer uses are permitted at run-time.
Since stack corruptions have become less important in recent years while 
heap corruptions are still very common and often
critical~\cite{miller2019trends},
\Name focuses on heap safety. Nonetheless, \Name offers not only memory safety
for heap objects, but also protects its safe heap against other accesses, \eg
using stack-based or external pointers.

Compared\,to\,other\,approaches\,of\,similar\,preciseness\,\cite{ABS94,JMG+02,NMW02,NZMZ09,BMCP18},
\Name's constant maximum memory overhead of only $2$ MB on systems with $17$ unused pointer
bits ($2^{17}$ table entries with $16$ bytes each) is almost negligible.
\Name is fully compatible with uninstrumented code, enabling the safe use and
creation of external pointers in \Name-instrumented code without manual code
modifications.

In summary, we make the following contributions:
\begin{itemize}
	\item We propose \Name, a memory-efficient safe heap using code instrumentation
		for spatial and temporal memory safety.
	\item We present an LLVM-based prototype implementation
		protecting the heap of C/\Cpp applications in a Linux runtime
		environment for the \archx and \archarm
		architectures.\footnote{Source code available under: https://github.com/Fraunhofer-AISEC/mesh}
	\item We show the feasibility of our approach in a detailed
		evaluation, measuring and comparing \Name's performance and memory
		overhead to existing solutions.
\end{itemize}
The remainder of the paper is organized as follows. In
\Cref{sec:goals}, we define design goals for \Name. \Cref{sec:design} details
\Name's actual design based on our design goals.
We describe our prototype in \Cref{sec:implementation} before
discussing evaluation results in \Cref{sec:eval}. Finally, we examine related work in
\Cref{sec:relatedwork} and conclude in \Cref{sec:conclusion}.

\section{Design Goals}\label{sec:goals}

For the design of \Name, we assume an attacker who can perform memory corruption
attacks on objects in arbitrary data regions of a program. Further, we assume
that the attacker cannot corrupt the code itself (typically
guaranteed by non-writable text segments) or modify the program's data regions from outside
the instrumented code, \eg using special hardware access.

Based on these assumptions, we define six design goals for \Name to ensure
heap memory safety for C/\Cpp while
maintaining high compatibility and memory efficiency.
Our first three design goals aim to \emph{ensure memory safety} for the \Name
heap: 

\begin{itemize}
\item[\goalone] \textbf{Detect temporal bugs:} Detect the usage of heap pointers to
	heap objects that are no longer allocated (\ie dangling pointers).
\item[\goaltwo] \textbf{Detect spatial bugs:} Detect any access (read/write)
	using a heap pointer pointing outside the allocation bounds of its object (\ie overflows or underflows) with byte precision.
\item[\goalthree] \textbf{Detect unprotected pointer accesses:} Detect
	accesses to protected objects using unprotected pointers (\eg stack or
	external pointers).
\end{itemize}
The next design goals target \Name's \emph{compatibility}. Our defense
mechanism must not change the functionality of the software it protects and
developers should not have to be aware of the underlying protection. In other
words, \Name has to work on any existing C/\Cpp code without requiring
modifications. We formalize these goals as follows:
\begin{itemize}
\item[\goalfour] \textbf{Support standard C/\Cpp:} Every language feature in the C/\Cpp standards must be supported. Moreover, language features should not be limited or extended.
\item[\goalfive] \textbf{Support unprotected code:} Code that is instrumented must be compatible with uninstrumented code, \ie calling external functions or using pointers created in external code must not break the functionality of the program.
\end{itemize}
Finally, the last design goal aims to enable \Name's \emph{application in memory-constrained environments}:
\begin{itemize}
\item[\goalsix] \textbf{Impose virtually no memory overhead:} The memory overhead imposed on the protected software must be constant and insignificant in the software's overall memory consumption.
\end{itemize}

\section{\Name Design}\label{sec:design}
In C and \Cpp, pointers are simple types that hold addresses without
any information about the memory object pointed to. 
Hence, programming errors involving pointers can easily lead
to violations of temporal or spatial memory safety.
To ensure memory safety, pointers must be augmented with additional
\emph{metadata} that can be used to check if an access is legal at run-time. 
\Name keeps its metadata in a table disjoint from pointers and objects,
and uses pointer tagging for linking pointers to metadata.

In the following, we first introduce the \Name metadata structure. Then, we discuss the management and placement of our
safe heap. Finally, we discuss how the \Name heap achieves our goals in
terms of memory safety and compatibility.

\subsection{Objects and Metadata}\label{subsec:lifecycle}
To discuss details of \Name's design, we first identify
all operations on objects that must be considered when implementing memory safety. The life-cycle of heap
objects in C/\Cpp
has three phases:
\begin{description}
\item[Allocation.] The allocation of objects in the heap is done by calling
	the corresponding functions (\eg \textit{malloc} or \textit{new}).
	The allocator then reserves enough bytes of memory for the object 
	and returns a pointer to the start of the object.
\item[Access.] Read or write accesses must happen
	during this phase. Accesses are only legal on allocated objects and
	must be within object bounds. Otherwise, they present spatial memory bugs, \eg \emph{buffer overflows and underflows}.
\item[Deallocation.] The deallocation of heap objects is done manually using the
	respective allocator functions (\eg \textit{free}
	or \textit{delete}). Only one deallocation per allocated
  object is allowed. Deallocating an object more than once, a
  \emph{double-free} bug, or accessing a deallocated object, a
	\emph{use-after-free} bug, can directly lead to memory corruptions or information leaks.
\end{description}
\subsubsection{\Name Metadata Structure}

To ensure memory safety, \Name requires metadata to describe the spatial and 
temporal characteristics of objects. This metadata can be represented as follows:
\begin{equation*}
    m_{obj}:=\Big(\big(lower\_bound_{obj},upper\_bound_{obj}\big),validity\_flag_{obj}\Big)
\end{equation*}
For spatial memory safety, all pointer \emph{accesses} must be within the
bounds of the object they are pointing to. Memory associated with
objects is contiguous, starting from a lower bound and ending at an upper
bound. 
These bounds are part of the object metadata and are used for checking
each pointer access.
For ensuring temporal safety, all pointer \emph{accesses} must happen on objects that 
have been allocated, but not yet deallocated. Therefore, the metadata must 
contain information about the temporal validity of the object. For this 
purpose, a metadata entry contains a validity flag, which 
is set during allocation and cleared as soon as the object is 
deallocated. 

\Name stores all metadata disjoint from pointers and objects in 
the \Name table, as illustrated in \Cref{fig:metadata}. Each row in the table uniquely corresponds to
one object $obj$ containing its metadata $m_{obj}$. The rows are indexed using
the global \Name table index $I$. \Cref{fig:metadata} additionally shows a
pointer $c$ to object $obj$ at address $N$. During the \emph{allocation} of $obj$, a new entry is
generated at the current index $I$. The entry contains the object's
base address $N$ as the $lower\_bound_{obj}$ and the result from adding the
size of $obj$ to its base address as the $upper\_bound_{obj}$. The
$validity\_flag_{obj}$ is also set and cleared once the object is \emph{deallocated}.

\subsubsection{\Name Metadata Access}

\Name uses unused pointer bits to store an index into
the \Name table, linking a pointer to its corresponding object's metadata entry.
\begin{figure}[tbp]
  \centering
  \includegraphics[width=1.05\columnwidth]{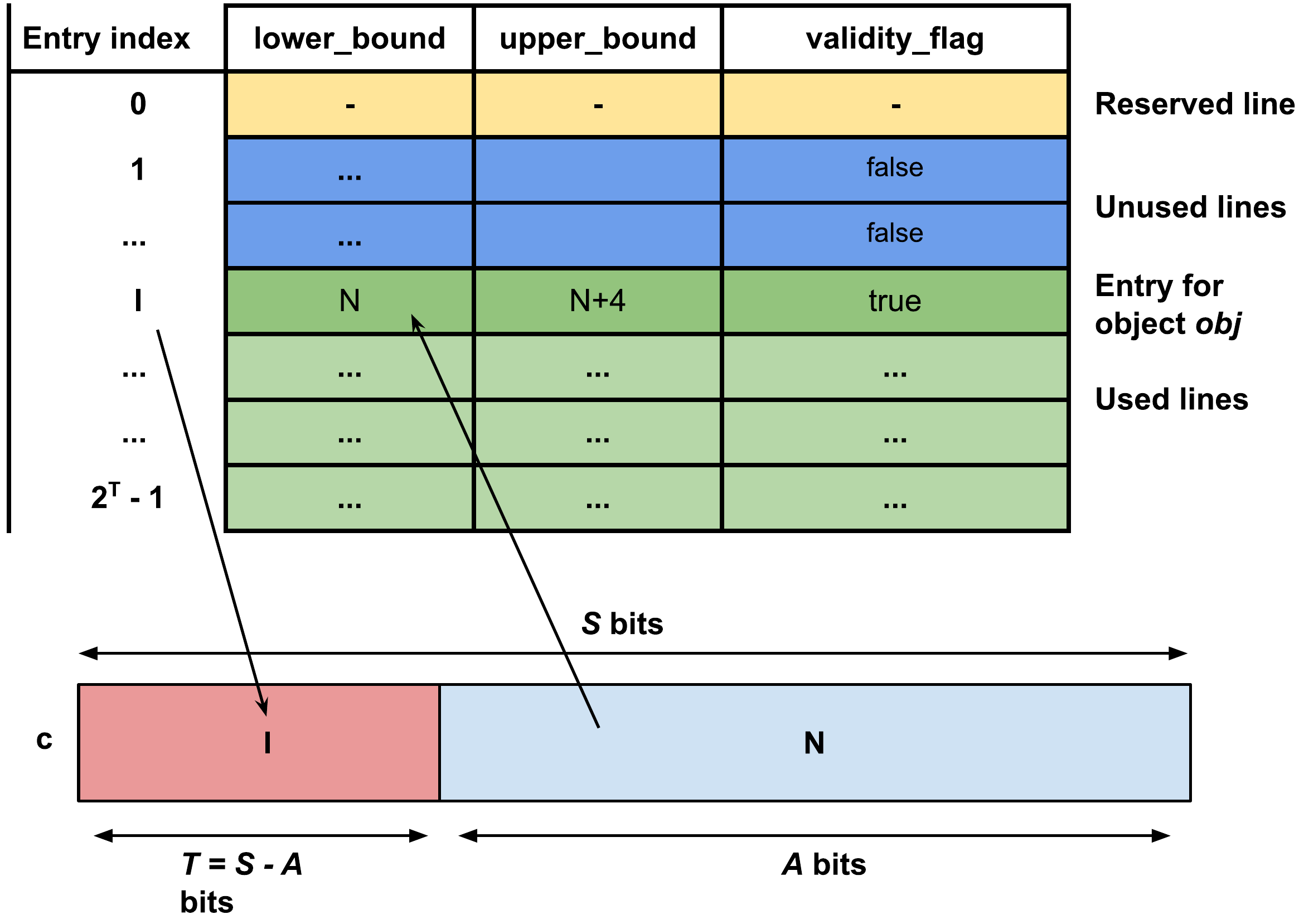}
  \caption{\Name table after a 4-byte allocation at address N, and its resulting tagged pointer}
  \label{fig:metadata}
\end{figure}
This is illustrated in \Cref{fig:metadata}, showing a system in which $T$ pointer bits are unused.
By storing the index $I$ in those
unused pointer bits, \Name links the pointer $c$ to its object $obj$, represented by
entry $I$ in the \Name table. 
$T$ also determines the size of the \Name table (\ie $2^T$ possible entries),
limiting the number of objects that can be protected simultaneously.

In a sequence of allocations, $I$ always contains the \Name table index
used for the next object. At program start, $I$ is initialized to the maximum value of
$2^T - 1$ and the \Name table is initialized to contain only invalid entries.
During program execution, $I$ is decremented
after each allocation. If $I$ reaches its minimum value of $1$, 
\Name will either wrap around $I$ and start re-using table slots freed in the
meantime or will terminate on the next allocation.
This is discussed further in \Cref{subsec:limitations}.

\subsection{Heap Separation}
\label{sec:heap_separation}

For \Name to effectively protect its heap objects, we must not only guarantee
their memory safety by maintaining the \Name table and tagging pointers, but
also secure access to those objects and the \Name table from unprotected, \ie
\emph{untagged}, pointers.
Unprotected pointers might origin from within the program itself, \ie from
global, stack, or memory-mapped memory,
or from a call to an external function that returns a pointer to an object allocated in the unprotected heap or memory-mapped memory of the program.

To achieve this protection, we separate the \Name table and all
\Name-protected heap objects from other data segments in a dedicated memory segment, the \emph{safe heap}.
The safe heap is managed by its own allocator, enabling programs to allocate protected objects with tagged pointers on the safe heap, while simultaneously supporting the allocation of unprotected objects using the conventional allocator and generating untagged, \ie unmodified, pointers.
To protect safe heap objects from accesses through untagged pointers, we
prevent any access to the safe heap using an untagged pointer. %into the safe heap.
As a consequence, the \Name index $I=0$ is reserved for untagged pointers and cannot be used to store \Name metadata.

\subsection{Memory Safety Enforcement}\label{subsec:pointers_op}

To show how \Name enforces memory safety, we examine possible operations on
pointers. 
A pointer is generated whenever a new heap object is allocated and its memory address is \emph{assigned} to the pointer.
Additionally, a pointer may also be \emph{derived} from another, already assigned pointer.
During the object's life span, the pointer can be \emph{compared} to other pointers or \emph{dereferenced} to access the underlying object.
Once the object is deallocated, the pointer is \emph{invalidated} and no further use should be permitted.
When a pointer is derived from another pointer, an implicit invalidation of the overwritten pointer occurs.
To achieve our design goals from \Cref{sec:goals}, the \Name design must
consider all five pointer operations.

\subsubsection{Assignment}\label{subsubsec:pointers_op_assignment}

With \Name, objects can be allocated in one of two coexisting memory areas:
the \Name-protected safe heap and the remaining memory, including allocations
on the stack, the normal heap, or in memory-mapped memory.
In general, \Name automatically redirects all heap allocations inside the program to the safe heap, while other allocations remain unaffected.
In particular, the latter includes allocations in external, uninstrumented code, such as linked libraries, which are unchanged by \Name, and still reserve memory in the normal heap and return untagged pointers.
Using this separation, \Name does not interfere with any C/\Cpp feature nor restricts compatibility with uninstrumented code, fulfilling our design goals \goalfour and \goalfive.

For allocations on the safe heap, \Name must generate a new metadata entry and
tag the new pointer with its \Name table index.
For this, we wrap the safe heap allocator (\texttt{safe\_malloc}) within a dedicated metadata generation routine, as shown in \Cref{alg:allocation}.
The routine takes the allocation size as input and first tries to allocate the object through the safe heap allocator.
If the allocation fails, the routine aborts and returns \texttt{NULL} letting the application handle the failed allocation.
Next, the routine derives the table index $i$ for the newly allocated
object from the global \Name table index~$I$.
The operations on $I$ are arranged in a way that the allocation algorithm can be made thread-safe as discussed in~\Cref{sec:multithreading}.
Finally, the routine initializes the bounds and validity metadata, and returns the tagged address as the final pointer to the object.

\begin{algorithm}[tb]
\caption{Metadata generation during allocation\label{alg:allocation}}
\begin{algorithmic}[1]
\Globals
\State $Table$: \Name table
\State $I$: \Name table index
\EndGlobals
\Procedure{mesh\_malloc}{$size$}
\State $address\gets$ \Call{safe\_malloc}{$size$} \Comment{Allocate safe heap object}
\If{$pointer=NULL$}
\State \Return{$NULL$}
\EndIf
\If{$I=0$} \Comment{Check for end of \Name table}
\State \Fail{Metadata exhaustion}
\EndIf
\State $i\gets I$ \Comment{Derive new index}
\State $I\gets I-1$ \Comment{Decrement index}
\State $Table[i].lower\_bound\gets address$ \Comment{Set spatial metadata}
\State $Table[i].upper\_bound\gets address+size$
\State $Table[i].validity\_flag\gets true$ \Comment{Set temporal metadata}
\State $address\gets i\ |\ address$ \Comment{Tag pointer}
\State \Return{$address$}
\EndProcedure
\end{algorithmic}
\end{algorithm}

\subsubsection{Derivation}\label{subsubsec:pointers_op_derivation}

Pointers can be derived from other pointers by copying or performing pointer
arithmetic. 
Pointers can also be cast to different types, changing the way they are dereferenced.
Since the derived pointer must retain the same restrictions as the
pointer it derived from, \Name
must propagate metadata from source to destination pointers.

As \Name uses tagged pointers,
metadata is an implicit part of the value used during the pointer derivation,
trivially achieving design goal \goalfour without instrumentation.
For copying and casting, the tag is implicitly copied with the pointer value.
Similarly, pointer arithmetic directly performs calculations on the tagged
pointer and generates a new pointer containing the same tag. As desired,
untagged pointers remain untagged through pointer derivation.
During pointer arithmetic, address values might theoretically overflow into
the \Name tag bits, potentially corrupting an existing tag or creating a tag
in an untagged pointer.
It is very unlikely that the result of such overflows 
is usable, since the pointer value must match the bounds associated with the
corrupted or crafted tag. 

\subsubsection{Comparison}\label{subsubsec:pointers_op_comparison}

In C and \Cpp, pointers of the same object can be compared for equality or relative ordering.
With \Name, pointers of the same object share the same tag, leaving comparisons for equality or relative ordering intact without instrumentation.
C and \Cpp do not define comparing pointers of different objects and comparing pointers with different tags in \Name might yield different results than comparing untagged pointers without \Name.

\subsubsection{Dereferencing}\label{subsubsec:pointers_op_dereferencing}

\begin{algorithm}[t]
\caption{Temporal safety check}\label{alg:temporal_checks}
\begin{algorithmic}[1]
\Globals
\State $Table$: \Name table
\EndGlobals
\Procedure{temporal\_safety\_check}{$pointer$}
\State $tag\ |\ address \gets pointer$ \Comment{Split pointer}
\If{$tag=0$} \Comment{No check for untagged pointers}
\State \Return{}
\EndIf
\Statex\LeftComment{1}{Perform temporal check}
\If{$Table[tag].validity\_flag=false$}
\State \Fail{use-after-free detected}
\EndIf
\EndProcedure
\end{algorithmic}
\end{algorithm}

An object is accessed by dereferencing
the corresponding pointer.
For safe heap pointers, accesses must only be allowed within the bounds of 
the object and during the life span of the object, \ie while it is still allocated.
Both constraints are enforced in \Name by calling the temporal and
spatial safety check routines shown in
\Cref{alg:spatial_checks,,alg:temporal_checks} before dereferencing a pointer.
Both routines first separate the tag from the checked pointer and then use the
metadata from the \Name table entry indexed by the tag to verify the legality
of the access. A temporal memory safety violation is detected if the pointer's
object has been deallocated, indicated by an unset validity bit. A spatial
violation is
detected if the access is outside the object's bounds.
The spatial check takes the pointer type's size into account, also detecting
overflows that occur due to casting pointers to larger sizes.
The \Name table is initialized with unset validity flags and invalid bounds,
so that tag bits indexing an unused metadata entry will always lead to a
violation.
Pointers not derived from an allocation in the safe heap should never be
allowed to access the safe heap. Hence, for those untagged pointers, the
spatial safety check must additionally ensure that access to the
safe heap memory is prohibited.

After the temporal and spatial validity of a pointer has been verified, the pointer can be dereferenced.
However, because \Name uses tagged pointers, we cannot directly take the pointer's value to access the underlying memory.
Hence, \Name \emph{strips} pointers, \ie removes their
tag bits, before dereferencing.  % from its value.

In summary, by calling both the temporal and spatial safety check routines
before dereferencing a pointer, \Name fulfills the design goals \goalone,
\goaltwo, and \goalthree.
In addition, by stripping the pointer after validation but before dereferencing, design goal \goalfour is also met.

\begin{algorithm}[tb]
\caption{Spatial safety check}\label{alg:spatial_checks}
\begin{algorithmic}[1]
\Globals
\State $Table$: \Name table
\EndGlobals
\Procedure{spatial\_safety\_check}{$pointer$}
\State $tag\ |\ address \gets pointer$ \Comment{Split pointer}
\If{$tag=0$}
\If{$address \in \textrm{safe heap}$}
\State \Fail{Illegal access to safe heap detected}
\Else \State \Return{} 
\EndIf
\EndIf
\Statex\LeftComment{1}{Perform spatial checks}
\If{$address<Table[tag].lower\_bound$}
\State \Fail{Buffer underflow detected}
\EndIf
\State $size\gets$ \Call{sizeof}{\Call{typeof}{$pointer$}}
\If{$(address+size)\geq Table[tag].upper\_bound$}
\State \Fail{Buffer overflow detected}
\EndIf
\EndProcedure
\end{algorithmic}
\end{algorithm}

\subsubsection{Invalidation}\label{subsubsec:pointers_op_invalidation}

At the end of their lifetime, safe heap objects are deallocated using our dedicated allocator,
while objects on the normal heap are freed as regular.
To prohibit any further access to a deallocated object in the safe heap and to
achieve goal \goalone, \Name must invalidate the corresponding metadata entry.
Additionally, \Name must ensure that normal heap objects can still be freed
using untagged pointers and, to achieve goal \goalthree, that this
does not affect the safe heap.
Similarly to safe heap allocations, deallocations are realized by wrapping the safe
heap deallocator (\texttt{safe\_free}) within a metadata invalidation
routine.

The metadata invalidation routine, shown in
\Cref{alg:deallocation}, takes the pointer to be freed as input.
Typically, heap objects allocated by external, uninstrumented code are also
deallocated externally. Nonetheless, since this is not always the case, the
routine must be able to deallocate normal heap objects with
untagged pointers.
If the given pointer is untagged, the corresponding object is freed using the
system-provided heap deallocation function, after checking that the pointer
does not point into the safe heap region.
For tagged pointers, the routine performs a temporal safety check before
invalidating and freeing the object.

\begin{algorithm}[bt]
\caption{Metadata invalidation during deallocation}\label{alg:deallocation}
\begin{algorithmic}[1]
\Globals
\State $Table$: \Name table
\EndGlobals
\Procedure{mesh\_free}{$pointer$}
\State $tag\ |\ address\gets pointer$ \Comment{Split pointer}
\If{$tag = 0$}
\If{$address \in \textrm{safe heap}$}
\State \Fail{Illegal free in safe heap detected}
\Else 
\State \Call{free}{$address$} \Comment{Deallocate normal heap object}
\State \Return{}
\EndIf
\EndIf
\Statex\LeftComment{1}{Perform temporal check}
\If{$Table[tag].validity\_flag=false$} 
\State \Fail{double-free detected}
\EndIf
\State $Table[tag].lower\_bound\gets 0$ \Comment{Invalidate entry}
\State $Table[tag].upper\_bound\gets 0$
\State $Table[tag].validity\_flag\gets false$
\State \Call{safe\_free}{$address$} \Comment{Deallocate safe heap object}
\EndProcedure
\end{algorithmic}
\end{algorithm}

\subsection{Compatibility}
\label{sec:compatibility}

As discussed in the previous section, \Name's pointer tagging naturally interferes with C/\Cpp
language assumptions. However, because \Name strips pointers before
dereferencing or calling any deallocation functions, design goal \goalfour is
still fulfilled.
For implementing \Name, the only essential requirement are unused bits in
pointers to store the \Name table index $I$. \Name does not make any other
assumptions about the underlying platform implementation nor modifies it in any
way. In particular, while \Name moves heap allocations into the safe heap, it does
not change the behavior or semantics of a program's normal heap and its memory management.
Further, pointers are not \enquote{fattened} by \Name using pointer structures
and pointers to globals, stack objects, or other unprotected memory are left untouched.
Moreover, the most notable sources of language
incompatibility for memory safety mechanisms are modifications to function
call conventions, casts between different pointer types, or casts between pointers
and integers. \Name handles these cases by encoding the index to
the metadata into the value of the pointer itself: Since pointers are passed
to functions by value, their tags are passed implicitly. Likewise, casts are done
by value, only having their interpretation changed. 

For compatibility with external code, we have to
consider untagged pointers that originate from uninstrumented code, and
tagged pointers that are passed to uninstrumented code.
Since the normal heap is still available alongside the safe heap, uninstrumented code can still
use the system-provided heap management functions, allocating
and deallocating objects with untagged pointers.
If those untagged pointers are returned to \Name-instrumented code, \Name is able to use them as-is
without breaking any functionality and without compromising the security of
the safe heap, as those untagged pointers can never access the safe heap.

When passing tagged pointers as arguments to external
functions, the pointers are checked and stripped by \Name before issuing the
function call, as they are unusable for uninstrumented code otherwise.
For functions returning one of its pointer arguments, \Name additionally wraps the call to store the stripped tag temporarily and reapply it to the returned pointer argument after the function returned.
To identify functions external to \Name-instrumented code, we perform
instrumentation during \gls{lto}. \gls{lto} generates an inter-modular
representation of the entire application, enabling our instrumentation to
detect truly external functions that will not be instrumented. As a
result, \Name achieves complete compatibility with any external, uninstrumented
library, and therefore meets design goal \goalfive.

\subsection{\Name Limitations}\label{subsec:limitations}

The limitations of \Name result from the number of unused bits in a pointer. As those bits store the \Name table index $I$,
they directly determine the maximum number of objects that can be represented
in the \Name table.

\subsubsection{Heap-Only Protection}

In general, we designed \Name to avoid the complete filling of its table.
As a consequence, we are bound to protecting a maximum of $2^T-1$ objects, when $T$ pointer bits are unused.
In practice, as shown in \Cref{sec:eval_allocations}, restricting the \Name protection to the heap
alleviates this problem, as there are generally much fewer, but more long-lived
heap objects than stack objects in typical programs.

However, limiting the \Name protection to the heap only, leaves stack objects unprotected and stack pointers untagged.
An unprotected stack does not pose a direct security threat to \Name, as untagged pointers are prohibited from accessing the safe heap.
But, with a stack vulnerability present, an attacker might be able to craft a tagged pointer to access the safe heap regardless.
To circumvent this possibility, it is possible to combine \Name with a separate stack protection mechanism, such as Safe\-Stack \cite{KSP+14}.
With Safe\-Stack, potentially exploitable stack objects are moved to \Name's safe heap, while provable safe objects remain on the stack.
Consequently, stack vulnerabilities no longer pose a security threat as well.

\subsubsection{Architecture Support}

To reduce the limiting effects of unused pointer bits available for storing the \Name table index $I$, \Name is best supported on 64-bit architectures.
On our target architectures \archx and \archarm, \Name natively supports
storage of at least $2^{17}-1$ and $2^{16}-1$ metadata entries, respectively.
Most applications---especially in a memory-constrained environment---only use
up to a few thousand heap objects per process (as shown
in more detail in \Cref{sec:eval_allocations}). Moreover, 
by modifying the kernel's virtual
address space size (\ie decreasing the number of used address bits in a
pointer) or by restricting the heap memory space on an application level, the
bits available for storing $I$ can be increased.

\subsubsection{\Name Table Wrap-Around}

If more than the supported $2^T - 1$ allocations are required by an
application, once the \Name table index $I$ reaches its minimum value of $1$,
\Name can optionally perform a wrap-around of $I$ and reuse entries of 
objects freed in the meantime. If enabled, this wrap-around is
an extension to \Name's metadata generation routine (see \Cref{alg:allocation})
that searches for entries in the \Name table whose validity flag is unset. Such
an entry is then reused by the routine to store the metadata of the newly
allocated object. While this wrap-around avoids the limitation on the total
number of allocations, \Name is still limited to support at most $2^T - 1$
alive safe heap objects at the same time. 
The limit on simultaneously alive objects---with or
without the optional wrap-around---is similar in nature to other common
resource exhaustion problems, such as stack overflows or heap exhaustion.
However, the optional wrap-around degrades \Name's temporal safety guarantees from
precise to probabilistic: A \emph{dangling} pointer to an already deallocated object might get usable
again when the metadata entry corresponding to its tag gets reused due to the
wrap-around mechanism. 
Nevertheless, for rogue memory accesses to succeed with such dangling pointers, the
bounds of the new object have to overlap with the bounds of the past object, as
otherwise a spatial safety violation is detected. Considering this, we can
assume that most memory safety violations are still detected and \Name can,
with its wrap-around mechanism, offer an attractive trade-off for larger
applications.

\section{Implementation}
\label{sec:implementation}

To show the feasibility of \Name, we implemented a prototype for the \archx
and \archarm architectures as an extension to the LLVM compiler framework
(version 11).
Our prototype currently supports the ELF binary format for Linux runtime environments.

As discussed in \Cref{subsec:pointers_op}, \Name has to handle pointer assignments, dereferences, and invalidations to enforce the memory safety of heap objects.
Our prototype provides a dedicated runtime support library for assigning and invalidating pointers, and directly instruments \emph{load} and \emph{store} instructions in LLVM's \gls{ir} with the temporal and spatial safety checks for pointer dereferences.
Additionally, our prototype optimizes performance by removing checks statically proven to be safe and ensures that all changes to the program are thread-safe.

\subsection{Runtime Support Library}

We built our prototype's runtime support based on the LLVM compiler runtime (\texttt{compiler-rt}) library.
The runtime support initializes and manages the safe heap, maintains the \Name table, and handles memory safety violations.

The runtime support adds a constructor to protected binaries, which is executed by the dynamic linker at load-time.
This constructor first allocates $1/8$ of available memory for the safe heap using \texttt{mmap}.
Within the safe heap, it then allocates the \Name table and initializes the entire table to invalid memory addresses (\ie the value
\texttt{0xFF...FF} for both columns \emph{lower\_bound} and \emph{upper\_bound}).
As discussed in \Cref{subsec:lifecycle}, the \Name table has a constant size of $2^T$ entries, depending on the architecture and its number of unused bits $T$ in a pointer.
The constructor initializes the \Name table index $I$ to the table's uppermost entry, \ie $I=2^T-1$.
The index is also allocated in the safe heap, making it solely accessible to our metadata handling routines.
The \Name table and its index are independent for each process so that there are no inter-process conflicts.

The runtime support also provides our dedicated safe heap allocator and routines wrapping the allocator functions to handle \Name metadata.
The safe heap allocator is based on a simple heap allocator implementation\footnote{\url{https://github.com/CCareaga/heap_allocator}}
modified by us to be thread-safe and support 64-bit systems.
The implementation of the metadata routines, as described in \Cref{subsec:pointers_op}, straightforwardly follows
\Cref{alg:allocation} for allocating and \Cref{alg:deallocation} for deallocating.
The \Name instrumentation detects calls to the GNU C and \Cpp standard library's allocation and deallocation functions and
redirects them to our routines.\footnote{\Name provides custom routines for \texttt{malloc}, \texttt{free}, \texttt{new}, \texttt{delete}, \texttt{calloc}, \texttt{realloc}, and \texttt{memalign}.}
In addition, if enabled through a compiler flag, the allocation routines also perform the optional \Name table
wrap-around, as detailed in \Cref{subsec:limitations}.

\subsection{IR Instrumentation}

To enforce temporal and spatial memory safety, our prototype has to instrument
each load and store instruction. 
Because loads and stores are very frequent, we use an optimized check that combines the temporal and spatial checks (\Cref{alg:temporal_checks} and \Cref{alg:spatial_checks}) presented in \Cref{subsec:pointers_op}.
This optimized check routine, shown in \Cref{alg:checks}, requires a specific
placement for the safe heap and an adaption of the \Name table, both
described in the following.

\begin{algorithm}[tb]
\caption{Spatial and temporal safety check}\label{alg:checks}
\begin{algorithmic}[1]
\Globals
\State $Table$: \Name table
\EndGlobals
\Procedure{safety\_check}{$pointer$}
\State $tag\ |\ address\gets pointer$ \Comment{Split pointer}\label{alg:checks:split_pointer}
\If{$tag=0$}\label{alg:checks:safe_heap_check_begin}
\Statex \LeftComment{2}{Optimized access to safe heap check}
\If{$address \leq \textrm{safe heap upper bound}$}\label{alg:checks:safe_heap_bound}
\State \Fail{Illegal access to safe heap detected}
\Else \State \Return{}\label{alg:checks:safe_heap_check_end}
\EndIf
\EndIf
\Statex \LeftComment{1}{Combined temporal and lower bound spatial check}
\If{$address<Table[tag].lower\_bound$}\label{alg:checks:lower_bound}
\State \Fail{use-after-free or buffer underflow detected}
\EndIf
\Statex \LeftComment{1}{Upper bound spatial check}
\State $size\gets$ \Call{sizeof}{\Call{typeof}{$pointer$}}
\If{$(address+size)\geq Table[tag].upper\_bound$}\label{alg:checks:upper_bound}
\State \Fail{Buffer overflow detected}
\EndIf\label{alg:checks:end}
%\State \Return{}
\EndProcedure
\end{algorithmic}
\end{algorithm}

First, to protect the safe heap against spatial violations through untagged
pointers, our combined memory safety check must verify a pointer's address
against the bounds of the safe heap 
(lines \ref{alg:checks:safe_heap_check_begin} to
\ref{alg:checks:safe_heap_check_end}).
This verification can be implemented using a \emph{single} comparison by placing the safe heap at a low address in memory before \emph{any other} data segment.
Hence, we allocate our safe heap at the lowest possible memory page, \ie the second page on Linux, requiring us only to check if the address of untagged pointers is greater than our safe heap's upper bound.

Then, the combined memory safety check must verify the temporal and spatial validity of tagged pointers using the metadata stored in the \Name table (lines \ref{alg:checks:lower_bound} to \ref{alg:checks:end}).
This verification can be implemented using only \emph{two} comparisons by optimizing the \Name table to encode a heap object's validity flag within its lower bound:
If the lower bound has a value of \texttt{0xFF...FF}, \ie the highest though invalid memory address, the entire \Name table entry can be considered invalid and any access to the corresponding object is prohibited.
Hence, revisiting \Cref{fig:metadata}, since the column \emph{validity\_flag}
is not required by the implementation, we can slim the \Name table to the two columns \emph{lower\_bound} and \emph{upper\_bound}.
The resulting combined check is identical to the spatial safety check from
\Cref{alg:spatial_checks}, except that the check in
\cref{alg:checks:lower_bound} can now indicate both a temporal or spatial
violation.
Both types of violations can still be differentiated afterwards by examining the
lower bound: A lower bound of \texttt{0xFF...FF} indicates a use-after-free
while other values indicate a buffer underflow.

\subsection{Check Removal}

By default, the \Name prototype instruments every pointer dereference.
This also includes dereferences loading from or storing to objects
that are not allocated on the heap and also do not have compound types.
Even though pointers to such objects are not tagged, our instrumentation still
requires three additional instructions to verify their tags are zero.
Hence, in order to reduce the performance impact of handling untagged pointers, we apply a simple optimization to our instrumentation.

For every pointer dereference, we perform an intra-procedural analysis of pointer origins.
If we infer that a pointer originates from an allocation on the stack (\ie resulting from an
\emph{alloca} instruction) or from a global \gls{ir} variable, we omit the instrumentation
of the corresponding pointer dereference if 
the accessed address is the one allocated and not otherwise derived.
The optimization is very conservative, only omitting instrumentation if a pointer's origin can be determined reliably.
For example, checks for accesses in which the pointer is going through a
PHI node are only omitted if it can be guaranteed that \emph{all} PHI
sources are stack allocations which have not been further derived.
This ensures that the optimization does not result in \Name missing tagged
pointers.
In an example case protecting the nginx web server with \Name, we were able to reduce the number of
instrumented pointer dereferences by about 5\% using our optimization.

\subsection{Multithreading Support}
\label{sec:multithreading}

Our \Name prototype is fully compatible with user-level and kernel-level multithreading.
In the following, we discuss the considerations taken into account for supporting multithreading and present how \Name achieves thread-safety.

\subsubsection{Concurrent Allocations}
When allocating objects on the safe heap, unique \Name table entries are required to store the objects' metadata.
To allow for concurrent allocations, our runtime support library must ensure
that the \Name table index $I$ is not modified concurrently.
To this end, our heap allocation routines use a global mutex to provide atomic access to the
retrieval of a new tag (\ie the modification of the index).
Thread-safety for the actual memory allocation is guaranteed by our underlying
heap allocator itself, as our runtime support library merely wraps the heap allocation functions.

\subsubsection{Concurrent Deallocations}
Deallocating the same object from two different threads is an application-level issue and must be handled by the application programmer.
In other words, without application-level thread-safety, this behavior is undefined with or without \Name.
Pointers of different objects are associated with different tags, hence,
correspond to different \Name table entries so that accesses to invalidate the metadata do not collide.

\subsubsection{Concurrent Allocations and Deallocations}
For the default configuration of \Name, concurrent allocations and deallocations already handled, as our allocation routines are thread-safe and our deallocation routines do not modify the \Name table index $I$.
However, if the \Name table wrap-around (see \Cref{subsec:limitations}) is active, concurrent allocations and deallocations become problematic as unused \Name table entries---which might concurrently be invalidated---can be reused for new allocations.
To solve this problem, we protect the invalidation of a \Name table entry with the same global mutex that is used by the allocation routines, making all allocations and deallocations atomic in regard to each other.

\section{Evaluation}\label{sec:eval}

To evaluate \Name, we first compare its
memory overhead to those of similar memory safety
solutions (as discussed in \Cref{sec:relatedwork}). Then, we evaluate the performance of \Name
on an artificial benchmark and a widely used real-world  program.
This evaluation is performed on the \archx and \archarm architectures.
Finally, we count the number of heap allocations in a mainstream application to determine the maximum
number of objects alive at the same time, as well as the total number
of allocations performed during the lifetime of the program. We use this to validate
our assumption that the \Name table is large enough for small to mid-size programs that typically run on
resource-constrained devices.

\subsection{Memory Overhead}\label{sec:memory_overhead}

\begin{table}[tb]
\centering
\caption{Memory overhead of memory safety mechanisms}
\label{table:mem_overhead}
\begin{tabular}{l l} 
 \toprule
 \textbf{Defense Mechanism} & \textbf{Memory Overhead (approx.)} \\ [0.5ex] 
 \midrule
 \Name & $\leq$ 2 MB (\archx) / $\leq$ 1 MB (\archarm)* \\
 CUP & $\leq  32$  GB $(2^{31} \cdot 16)$*~\cite{BMCP18}\\
 ASan & $> 200\%$~\cite{SLR+19}\\
 LFP & 3--11\%**~\cite{DY16}\\
 SoftBound (hash table) & 87\%~\cite{NZMZ09} \\
 SoftBound (shadow)& 64\%~\cite{NZMZ09}\\  [1ex]
 \midrule
 \multicolumn{2}{l}{\text{*} Cannot be measured in percent because it is constant} \\
 \multicolumn{2}{l}{\text{**} Depending on precision} \\
 \bottomrule
\end{tabular}
\end{table}
\Cref{table:mem_overhead} shows a comparison of \Name's memory overhead with
the memory overheads incurred by other memory safety solutions. 
\Name's and CUP's~\cite{BMCP18} overheads are given as calculated constant maximums, while the others
are variable and measured at run-time.
The comparison shows that, except for programs with little memory usage,
\Name achieves a much smaller memory overhead than the other approaches.
Especially solutions based on shadow memory, namely
Softbound~\cite{NZMZ09} and \gls{asan}~\cite{SBPV12}, typically perform much
worse than the approaches based on pointer tagging, namely LFP~\cite{DY16},
CUP, and \Name. Although LFP achieves a very low memory overhead,
in contrast to \Name, it only provides spatial safety and does not detect
memory corruptions with byte-precision, as discussed further in
\Cref{sec:relatedwork}.

\subsection{Performance Overhead}

For evaluating \Name's performance, we chose 
CoreMark\footnote{\url{https://www.eembc.org/coremark/}}, a CPU benchmark
designed for embedded systems, 
as a synthetic test and the Nginx\footnote{\url{https://www.nginx.com/}}
web engine in conjunction with the ApacheBench HTTP
benchmark\footnote{\url{http://httpd.apache.org/docs/current/programs/ab.html}}
as a real-world application test.
To evaluate the performance with Nginx, we let ApacheBench generate
$10{,}000$ HTTP
requests measuring the average response time. We repeated each test five
times to exclude outside factors as much as possible.
With our two benchmarks, we compared \Name and
\gls{asan} \cite{SBPV12} against a baseline without instrumentation.
Since \Name is a heap-only protection, we
configured \gls{asan} to also protect only the heap.

\Cref{perfcomp} summarizes the results of our evaluation. It shows that 
for CoreMark, \Name is noticeably slower than \gls{asan}.
This is mainly due to the more complex checking required for the
byte-precision and the space-saving \Name table, which, in this case, has a
heavy impact as CoreMark involves a large number of memory accesses.
However, in case of Nginx's real-world application test, \Name outperforms
\gls{asan} and causes almost no overhead at all. We explain this difference
with the better caching properties of the small \Name table in comparison to
\gls{asan}'s shadow metadata.
While CoreMark repeatedly accesses the same memory regions, for which the
metadata can be cached well for both solutions, Nginx accesses
a larger variety of memory regions, making the checks hard to cache for \gls{asan},
while remaining easy for \Name.

Summarizing, depending on the test case, \Name provides performance that is
comparable to \gls{asan}. However, \Name provides an increased precision, as discussed further in
\Cref{sec:relatedwork}, and has a lower memory footprint, as shown in \Cref{sec:memory_overhead}.
\begin{table}[t]
\centering
\caption{Performance overhead of \Name and \gls{asan}}
\label{perfcomp}
\begin{tabular}{c c c} 
	\toprule
	\textbf{Program} & \textbf{\Name} & \textbf{\gls{asan} (Heap-only)} \\
 \midrule
 CoreMark for \archx & 170\% & 51\% \\
 CoreMark for \archarm & 111\% & 30\% \\
 Nginx for \archx & 5.6\% & 8.2\% \\
 Nginx for \archarm & 3.1\% & 4.6\% \\
 \bottomrule
\end{tabular}
\end{table}

\subsection{Other Metrics}
\label{sec:eval_allocations}

Finally, to validate our assumption that the \Name table has enough entries for practical use,
we analyzed the number of allocated objects in Nginx and CoreMark. 
For Nginx, we counted the allocations for the same ApacheBench test procedure
used in the performance evaluation. Since Nginx is a multi-process
application and each process uses an independent \Name table, we simply took
the maximum number of allocations out of all processes.

The results show that the number of allocated heap objects varies substantially between
the two test programs. While CoreMark only allocated \emph{one} heap object over
its full execution in our test, Nginx allocated 5211 objects, of which only a maximum of
151 were alive at the same time.
In other words, only 4\% and 8\% of the maximum supported entries of the
\Name table were filled for \archx and \archarm, respectively. Furthermore,
only at most 0.1\% and 0.2\% of entires were used at the
same time.
The test results confirm that \Name can be used to effectively protect
real-world applications, especially in memory-constrained devices.

\begin{table*}[t]
	\footnotesize
\centering
\caption{Comparison of \Name and related memory safety solutions}
\label{table:overview_of_sanitizers}
\begin{tabular}{l c c c c c c c}
	\toprule
	Goal & \textbf{\Name} & \textbf{LFP}~\cite{DY16} & \textbf{\gls{asan}}~\cite{SBPV12} &
	\textbf{HWASan}~\cite{SSS+18}
	& \textbf{SoftBound/CETS}~\cite{NZMZ09,NZMZ10} & \textbf{CUP}~\cite{BMCP18} \\
 \midrule
 \goalone Detect temporal bugs & \textbf{Yes} & No & Imprecise & Probabilistic & \textbf{Yes} & Probabilistic\\
 \goaltwo Detect spatial bugs & \textbf{Yes} & Imprecise & Imprecise & Probabilistic & \textbf{Yes} & \textbf{Yes}\\
 \goalthree Detect unprotected pointer accesses & \textbf{Yes} & No & Imprecise & \textbf{Yes} & \textbf{Yes} & \textbf{Yes}\\
 \goalfour Support standard C/C++ & \textbf{Yes} & \textbf{Yes} & \textbf{Yes} & \textbf{Yes} & \textbf{Yes} & \textbf{Yes}\\
 \goalfive Support unprotected code & \textbf{Yes} & \textbf{Yes} & \textbf{Yes} & \textbf{Yes} & No & No\\
 \goalsix Impose low memory overhead & \textbf{Yes} & \textbf{Yes} & No & No & No & No\\
 \bottomrule
\end{tabular}
\end{table*}

\section{Related Work}\label{sec:relatedwork}

In the following, we analyze how \Name compares to other memory safety approaches.
We are mainly interested in solutions that meet at least some of the
\Name design goals presented in \Cref{sec:goals}.

\emph{Fat pointer} approaches~\cite{ABS94,JMG+02,NMW02}
store an object's metadata alongside its pointers (\eg using struct-like pointers).
Because metadata is stored directly with the protected pointers,
such approaches usually only tackle spatial memory safety and do not offer
temporal memory safety (\goalone): If an object
is deallocated, all its pointers must be found and invalidated,
which is not trivial without additional data structures.
Moreover, replacing normal pointers with fat pointers can cause incompatibilities
with language features and uninstrumented libraries,
thus failing to meet design goals \goalfour and \goalfive.
To tackle this problem, LFP \cite{DY16} encodes the metadata (still only spatial) inside the value of the pointer itself to protect the heap of 64-bit
systems. Similarly to \Name, LFP instruments LLVM \gls{ir} to tag pointers and associate them
with their object's spatial metadata.
But since LFP optimizes performance overheads by size-aligning objects
to a set of predefined sizes, it loses precision for its spatial checks
compared to \Name, as overflows up to the alignment size are not detected (\goaltwo).
In addition, LFP does not restrict access of non-heap and external pointers, failing to protect against unprotected pointers (\goalthree).
However, while not constant, LFP achieves a low memory overhead of 3-11\% (\goalsix).

Another approach to memory safety is the use of \emph{shadow memory}.
\emph{Location-based} solutions shadow a portion of the
addressable memory to store certain attributes (\eg accessible or
non-accessible) about the shadowed memory. Using this shadow, objects are
surrounded by \emph{red-zones} \cite{HJ91,SN05,BZ11,SBPV12} or interleaved by \emph{guard pages}
\cite{Linux-EF,Microsoft-PH,DMW17} to detect spatial memory violations. The most prominent location-based
solution is \gls{asan} \cite{SBPV12}, which shadows 8 byte blocks with an 8-bit
tag to insert red-zones between memory allocations. \gls{asan} performs
compile-time instrumentation of pointer dereferences to detect errors when
accessing the red-zones. 
\gls{asan} is not byte-precise (8-byte alignment) and only able to detect buffer under- and overflows, but not arbitrary reads and writes skipping red-zones.
Hence, unlike \Name, \gls{asan} only partially achieves spatial memory safety (\goaltwo and \goalthree).
Furthermore, and also in contrast to \Name, \gls{asan} only approximates temporal safety by delaying the reuse of freed memory (\goalone) and introduces a substantial memory overhead overhead of 237\% \cite{SBPV12}.
But, \gls{asan} supports standard C/\Cpp (\goalfour) and is compatible with uninstrumented code (\goalfive).

An alternative to \gls{asan} is \gls{hwasan}~\cite{SSS+18}, which aims to reduce memory and performance
overheads by using specific hardware features.
\gls{hwasan} implements a typical \emph{memory tagging} approach, in which
objects in memory and their pointers are tagged with tags that must match for
a memory access to be allowed. Similarly to \gls{asan}, \gls{hwasan} uses shadow
memory to tag objects, and similarly to \Name, uses unused
pointer bits to store its tags.
Since \gls{hwasan}, in its main form, relies on an \archarm-specific feature that allows the
processor to ignore the top eight bits in pointers, only 8-bit sized tags are used.
While not having to strip pointers before dereferencing significantly
increases \gls{hwasan}'s performance, its small tag size makes both spatial and temporal
checks probabilistic (\goalone and \goaltwo), with a chance of 0.39\% \cite{SSS+18} for missing bugs even
without an attacker targeting the mechanism. Furthermore, using shadow memory,
similarly to \gls{asan}, \gls{hwasan} is less memory-efficient than \Name
(\goalsix). However, \gls{hwasan} supports standard C/\Cpp language features
(\goalfour) and untagged pointers do not interfere with its instrumentation
(\goalfive) nor jeopardize the security of tagged objects (\goalthree).

Another group of approaches using shadow memory are \emph{identity-based} solutions, which
track object bounds for each pointer. Using these bounds, pointer dereferences
are precisely checked to detect spatial memory violations. These solutions
either use \emph{per-object bounds tracking} \cite{JK97,RL04,ACCH09,YPC+10}, where pointers to the
same object share the same bounds, or \emph{per-pointer bounds tracking}
\cite{PF97,NZMZ09,NZMZ10,BMCP18}, 
where each pointer tracks its own object bounds. Per-pointer
bounds tracking is usually more precise, as per-object bounds have to be aligned to
powers of two.
SoftBound \cite{NZMZ09}, the most prominent identity-based solution, uses
per-pointer bounds tracking for spatial memory safety (\goaltwo). Additionally, it ensures temporal memory safety with its CETS \cite{NZMZ10} extension that invalidates pointers on
object deallocation (\goalone). SoftBound/CETS supports standard C/\Cpp
(\goalfour), but is incompatible with pointers originating from uninstrumented
code (\goalfive), for which it aborts due to missing metadata (\goalthree).
SoftBound/CETS shadows 8 byte blocks (\ie pointers) with 32 byte of metadata. To reduce memory
overhead, a variant of SoftBound/ CETS uses a disjoint table to store metadata
instead of shadow memory. While this metadata store is similar to \Name,
instead of tagging pointers, SoftBound/CETS derives a unique hash from each
pointer to access its metadata. 
On average, with an overhead of 87\% for its
hash table and 64\% for its shadow memory \cite{NZMZ09},
SoftBound/ CETS has a significantly higher memory footprint than \Name~(\goalsix).

CUP~\cite{BMCP18} is a recent approach that is similar to
\Name design-wise and aims to achieve temporal and spatial
memory safety. As \Name, CUP uses a constant metadata table to facilitate
per-pointer bounds tracking. But in contrast to \Name, CUP focuses on completeness,
partly sacrificing memory-efficiency and modularity in comparison.
CUP also leverages the unused bits in pointers on 64-bit architectures to store a link into the
metadata table.
Pointers in CUP are completely replaced with an index into the metadata table
and the pointer's offset into the object. While this frees up additional index bits in
pointers, it comes at a performance loss, as, in comparison to \Name, additional steps are required to prepare a tagged pointer
for an actual memory access. CUP achieves probabilistic temporal and
precise spatial memory safety (\goalone and \goaltwo), but the
resulting CUP metadata table is very large (\goalsix).
Further, while CUP supports standard C/\Cpp language features
(\goalfour), by design, it does not strip pointers for
uninstrumented code and mandates the use of a modified \texttt{libc} library
that is capable of allocating protected objects and handling tagged pointers.
Hence, uninstrumented code is not able to allocate unprotected objects with
untagged pointers (\goalthree) nor able to dereference tagged pointers outside of \texttt{libc} (\goalfive).

\Cref{table:overview_of_sanitizers} summarizes \Name's comparison to 
other  approaches.
None of the other approaches is able to achieve all our design
goals, confirming the necessity for \Name as a memory-efficient and highly
compatible alternative for memory-safe heap solutions.

\section{Conclusion}
\label{sec:conclusion}
We presented \Name, a simple yet efficient safe heap design for C and \Cpp programs.
\Name's metadata store and pointer tagging mechanism ensure constant memory
overheads and metadata look-ups.
We showed the feasibility of our concept with a full LLVM-based
implementation for both major 64-bit architectures, \archx and \archarm.
Our practical evaluation using the Nginx web server and the CoreMark CPU benchmark
shows that while the performance overhead imposed by \Name is 
similar to those of existing memory safety solutions,
\Name typically requires magnitudes less memory, causing an insignificant and constant memory
overhead, such as $2$ MB on \archx.
Its performance and memory efficiency, together with its increased precision and complete
compatibility with external code make \Name a viable solution for heap
memory safety, especially in resource-constrained scenarios.

\begin{acks}
	This work was supported by the \grantsponsor{FhG}{Fraunhofer Internal
		Programs}{} under Grant No. \grantnum{FhG}{PREPARE 840 231}.
\end{acks}

\bibliographystyle{ACM-Reference-Format}
\bibliography{mesh2021}
%\printbibliography

\end{document}